\documentclass[12pt]{iopart}
\usepackage[utf8]{inputenc}
\usepackage[T1]{fontenc}
\usepackage[
 backend=biber,
 style=numeric,
 sorting=none
 ]{biblatex}
 \AtEveryBibitem{\clearfield{pages}}
\AtEveryBibitem{\clearfield{volume}}
\AtEveryBibitem{\clearfield{number}}
\AtEveryBibitem{\clearfield{url}}
\AtEveryBibitem{\clearfield{issn}}
\AtEveryBibitem{\clearfield{isbn}}
\AtEveryBibitem{\clearfield{publisher}}
\AtEveryBibitem{\clearfield{note}}
\AtEveryBibitem{\ifentrytype{article}
    {\clearfield{eprint}}
    {}}
\makeatletter
\newcommand{\mainmatter}{%
  \setcounter{footnote}{0}%
  \patchcmd{\@makefntext}{\fnsymbol}{\arabic}{}{}%
  \patchcmd{\@thefnmark}{\fnsymbol}{\arabic}{}{}%
  \def\@makefnmark{\textsuperscript{\arabic{footnote}}}%
}
\makeatother
\usepackage{iopams}
\usepackage{capt-of}
\expandafter\let\csname equation*\endcsname\relax

\expandafter\let\csname endequation*\endcsname\relax
\usepackage{hyperref}
\usepackage{amsmath}

\usepackage{physics}
\usepackage{nicefrac}
\usepackage{graphicx}
\usepackage{wrapfig}
\usepackage[normalem]{ulem}
\usepackage{tikz}
\usetikzlibrary{math}
\usetikzlibrary{positioning}

\usepackage{mathenv}
\usepackage{multirow}

\def\PBS[#1,#2]%
{
\draw[blue,line width=1pt] (#1,#2+0.15) -- (#1+0.15,#2) -- (#1,#2-0.15) -- (#1-0.15,#2) -- (#1,#2+0.15);
\draw[blue,line width=1pt] (#1,#2-0.15) -- (#1,#2+0.15);
}

\def\LLWP[#1,#2]%
{
\draw[red,line width=1pt] (#1-0.25,#2-0.15) -- (#1-0.15,#2-0.25);
}

\def\LUWP[#1,#2]%
{
\draw[red,line width=1pt] (#1-0.25,#2+0.15) -- (#1-0.15,#2+0.25);
}

\def\RUWP[#1,#2]%
{
\draw[red,line width=1pt] (#1+0.25,#2+0.15) -- (#1+0.15,#2+0.25);
}

\def\RLWP[#1,#2]%
{
\draw[red,line width=1pt] (#1+0.25,#2-0.15) -- (#1+0.15,#2-0.25);
}

\def\LDet[#1,#2]%
{
\draw[line width=1pt] (#1-0.65,#2+0.5) -- (#1-0.5,#2+0.65) -- (#1-0.65,#2+0.8) to[out=90+45,in=90+45, looseness=2] (#1-0.8,#2+0.65) -- (#1-0.65,#2+0.5);
\draw[line width=1pt] (#1-0.8,#2+0.8) -- (#1-1,#2+1);
}

\def\RDet[#1,#2]%
{
\draw[line width=1pt] (#1+0.65,#2+0.5) -- (#1+0.5,#2+0.65) -- (#1+0.65,#2+0.8) to[out=90-45,in=90-45, looseness=2] (#1+0.8,#2+0.65) -- (#1+0.65,#2+0.5);
\draw[line width=1pt] (#1+0.8,#2+0.8) -- (#1+1,#2+1);
}
\DeclareRobustCommand\ILPBS{\tikz \PBS[0,0];}
\DeclareRobustCommand\ILWP{\tikz{\draw[red,line width=1pt] (0.2,0.1) -- (0.341,0.1); \draw[white] (0.2,0.0) -- (0.341,0.0);}}
\DeclareRobustCommand\ILDet{\tikz{\begin{scope}[rotate=45] \LDet[0,0];\end{scope}}}

\begin{document}


\title[The Influence of Experimental Imperfections on Photonic GHZ State Generation]{The Influence of Experimental Imperfections on Photonic GHZ State Generation}
\author{Fabian Wiesner\textsuperscript{1,$\dagger$}, Helen M. Chrzanowski\textsuperscript{2,3}, Gregor Pieplow\textsuperscript{3}, \\Tim Schr\"oder\textsuperscript{3,4}, Anna Pappa\textsuperscript{1,5}, Janik Wolters\textsuperscript{2,6}\\[1em]
\footnotesize{\textsuperscript{1} Electrical Engineering and Computer Science Department, Technische Universit\"at Berlin, 10587 Berlin, Germany\\
\footnotesize\textsuperscript{2} Institute of Optical Sensor Systems, Deutsches Zentrum für Luft- und Raumfahrt e.V. (DLR), 12489 Berlin, Germany\\
\footnotesize\textsuperscript{3} Department of Physics, Humboldt-Universität zu Berlin, 12489 Berlin, Germany\\
\footnotesize\textsuperscript{4} Ferdinand-Braun-Institut, 12489 Berlin, Germany\\
\footnotesize\textsuperscript{5} Fraunhofer Institute for Open Communication Systems - FOKUS, 10589 Berlin, Germany\\
\footnotesize\textsuperscript{6} Institute of Optics and Atomic Physics, Technische Universit\"at Berlin, 10623 Berlin, Germany}\\[1em]}

\ead{\textsuperscript{$\dagger$}f.wiesner@tu-berlin.de}
\vspace{10pt}
\begin{indented}
\item[]June 2024
\end{indented}
\setlength{\mathindent}{0pt}
\begin{abstract}
While the advantages of photonic quantum computing, including direct compatibility with communication, are apparent, several imperfections such as loss and distinguishability presently limit actual implementations. These imperfections are unlikely to be completely eliminated, and it is therefore beneficial to investigate which of these are the most dominant and what is achievable under their presence. In this work, we provide an in-depth investigation of the influence of photon loss, multi-photon terms and photon distinguishability on the generation of photonic 3-partite GHZ states via established fusion protocols. 
We simulate the generation process for SPDC and solid-state-based single-photon sources using realistic parameters and show that different types of imperfections are dominant with respect to the fidelity and generation success probability. 
Our results indicate what are the dominant imperfections for the different photon sources and in which parameter regimes we can hope to implement photonic quantum computing in the near future.
\end{abstract}
\section{Introduction}\mainmatter
Greenberger-Horne-Zeilinger or in short GHZ states \cite{GHZ, ExpGHZ} have increasingly gained attention in recent years. GHZ states are multipartite maximally entangled states exhibiting correlations that can help perform specific tasks better than bipartite entanglement. One such example is quantum network routing, as studied in \cite{butterfly,Hahn2019}. Other applications of GHZ states in quantum communication and cryptography include secret sharing \cite{PhysRevA.59.1829,Pont24}, measurement-based quantum repeaters \cite{zwerger2012}, conference key agreement \cite{CKArev,ExpCKA} and anonymity \cite{ACKA, Thalacker_2021}, while schemes to distribute them on a network have already been proposed \cite{Li:23}. GHZ states are also a promising resource for quantum computing; one can, in theory, merge them to create larger resource states \cite{hilaire2023} that facilitate measurement-based quantum computing (MBQC) \cite{MBQC, jozsa2005introduction, omkar2022, bartolucci2023}. 
Especially for linear optical quantum computing  \cite{LOQCrev, Knill2001, varnava2008, Kieling_2007}, 3-partite GHZ states are a popular input resource to generate larger resource states \cite{Three_to_UQC, Why..., omkar2022} and are already sufficient to demonstrate single qubit rotations in an MBQC-protocol \cite{walther2005, prevedel2007}.
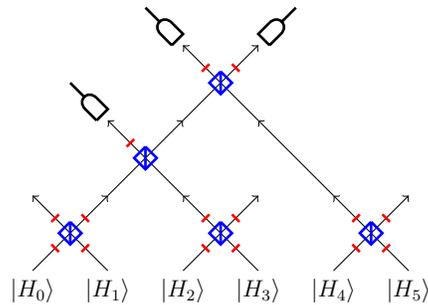
\begin{wrapfigure}{r}{0.45\textwidth}
    \centering
        \begin{tikzpicture}
            \draw[->] (0,0) -- (1,1) node[pos=0, below, scale=0.75] {$\ket{H_0}$};
            \draw[->] (1,1) -- (2,2);
            \draw[->] (2,2) -- (3,3);
            
            \draw[->] (1,0) -- (0,1) node[pos=0, below, scale=0.75] {$\ket{H_1}$};

            \draw[->] (2,0) -- (3,1) node[pos=0, below, scale=0.75] {$\ket{H_2}$};

            \draw[->] (3,0) -- (2,1) node[pos=0, below, scale=0.75] {$\ket{H_3}$};
            \draw[->] (2,1) -- (1,2);

            \draw[->] (4,0) -- (5,1) node[pos=0, below, scale=0.75] {$\ket{H_4}$};

            \draw[->] (5,0) -- (4,1) node[pos=0, below, scale=0.75] {$\ket{H_5}$};
            \draw[->] (4,1) -- (3,2);
            \draw[->] (3,2) -- (2,3);

            \PBS[0.5,0.5];
            \PBS[2.5,0.5];
            \PBS[4.5,0.5];
            \PBS[1.5,1.5];
            \PBS[2.5,2.5];

            \LUWP[0.5,0.5];
            \LLWP[0.5,0.5];
            \RUWP[0.5,0.5];
            \RLWP[0.5,0.5];

            \LUWP[2.5,0.5];
            \LLWP[2.5,0.5];
            \RUWP[2.5,0.5];
            \RLWP[2.5,0.5];

            \LUWP[4.5,0.5];
            \LLWP[4.5,0.5];
            \RUWP[4.5,0.5];
            \RLWP[4.5,0.5];

            \LUWP[1.5,1.5];

            \LUWP[2.5,2.5];
            \RUWP[2.5,2.5];

            \LDet[1.5,1.5];
            
            \LDet[2.5,2.5];
            \RDet[2.5,2.5];
        \end{tikzpicture}
        \caption{Circuit for heralded GHZ state preparation.``\ILPBS''\ is a polarizing beamsplitter, ``\ILWP''\ is a rotation of polarization by $45^{\circ}$ and ``\ILDet''\ is a polarization and number resolving detector. The circuit was originally proposed in \cite{varnava2008} and provides a success probability of $\nicefrac{1}{32}$, which is the probability for three single photon detections. In the ideal case, this indicates that the state is either $\nicefrac{\ket{H_0H_3H_5}+\ket{V_0V_3V_5}}{\sqrt{2}}$ or $\nicefrac{\ket{H_0H_3H_5}-\ket{V_0V_3V_5}}{\sqrt{2}}$ depending on the polarizations of the three measured photons.} 
        \label{fig:circuit}
\end{wrapfigure}
Photonic GHZ states can be created using probabilistic schemes based on spontaneous parametric down-conversion (SPDC) \cite{walther2005, prevedel2007, walther2004, zhong2018, wang2018}, and in deterministic schemes using single photon emitters, such as quantum dots \cite{schwartz2016, istrati2020} and single atoms \cite{thomas2022, thomas2024}. 

In this work, we focus on a very popular all-optical approach to the generation of GHZ states that is based on the interference and detection of single photons that heralds a 3-partite GHZ state. Despite recent advances in creating GHZ states using circuits like the one in Fig \ref{fig:circuit} \cite{cao2023photonic, chen2023heralded, Maring24}, the preparation of even relatively small (e.g. 3-partite) GHZ states with high fidelity remains an experimental challenge, which entails reliably preparing near-identical single photons, applying suitable interferometers with low loss and implementing photon-number resolved detection -- again with low loss. Unfortunately, small imperfections can have a catastrophic influence on the outcome. While it would be desirable to eliminate all imperfections, it could be more effective to focus on the more influential ones first and re-evaluate after technological advancements had a significant impact on the reduction of relevant errors. Such a realistic strategy poses the question of which imperfections influence the quality of the outcome the most -- and maybe even more importantly: What can we expect once we are in a low-error regime, and is it worth the required experimental effort, to address specific imperfections?

To answer these questions, we simulated the prototypical heralded ``GHZ state generation" circuit in figure \ref{fig:circuit} with realistic error models for photon distinguishability, losses and higher-order terms (emission events producing more than one photon), and computed the (post-selected) fidelity and success probability, i.e., the probability of accepting an outcome of the circuit per try. Our results indicate that losses and distinguishability are the two dominant sources of error in a realistic parameter range. This leads to a clear prescription for improving state-of-the-art setups: either one has to significantly reduce both of these imperfections or consider different, more error-resilient circuits. 

We present our results as follows: First, in the Methods section, we describe the simulation method, present the error models that are used and explain how we quantify the influence of the imperfections. In the next section, we examine the influence of the different errors for parameter regimes that are realistic for SPDC and solid-state single-photon emitters and we additionally investigate how a close-to-optimal choice of error parameters behaves. We conclude our paper with a discussion and open questions.
\section{Methods}
    \subsection{Simulation method}
    One can categorize recent simulations of linear optical circuits into two classes. The first one is based on the computation of the permanent of the unitary that is to be applied. For many settings, this is currently the fastest approach. An example of this class of simulations is the SLOS backend in Perceval \cite{parceval}. The second one does not consider the unitary as a whole but applies the individual parts of the circuits in consecutive order. While this is usually less performant, it allows the implementation of non-linearities such as distinguishability. One example is the stepper backend for Perceval \cite{parceval}.\\ 
    While the first class of simulations is of interest on its own, we focus on the second class. The reason for this choice is the error model; with indistinguishability, loss and many photon preparations, there are too many non-linearities for the permanent-based simulation method to work well. Another reason to choose the step-wise approach in the Fock representation\footnote{For an introduction to simulation of linear optical quantum computing in the Fock representation, we refer the reader to \cite{LOQCrev}.} is the reuse of already acquired data. We do that on two different levels.\\
    The higher level is about the decomposition of a mixed state and mixed operations. We consider a probabilistic mixture weighted with the probability for many photon creation events as the input. The operation we apply depends on the loss probabilities and is mixed with regard to these probabilities. We can change the probabilities of the mixture by first simulating the branches with the biggest impact (covering at least 98\% of the probability space represented by the set of events $E$). We can then use linearity of the fidelity and rearrange the output of the mixture from the separate branches using linearity of the fidelity to pure states:
    \begin{eqnarray}
        F\left(\rho,\ketbra{GHZ_3}{GHZ_3}\right) &=& \bra{GHZ_3}\rho\ket{GHZ_3} = \sum_{e\in E}Pr[e] \braket{GHZ_3}{\psi_e}\braket{\psi_e}{GHZ_3} \nonumber \\&=& \sum_{e\in E}Pr[e]\ F\left(\ketbra{\psi_e}{\psi_e},\ketbra{GHZ_3}{GHZ_3}\right),
    \end{eqnarray}
    where $\ket{\psi_e}$ is the prepared state if the event $e$ occurred and $Pr[e]$ is the corresponding probability. The same decomposition can be used for the success probability:
    \begin{eqnarray*}
        p_{\rm succ}(\rho) &=& \Tr\left(\pi_{\rm acc}\rho\right) = \sum_{e\in E}Pr[e]\ \Tr\left(\pi_{\rm acc}\ketbra{\psi_e}\right) = \sum_{e\in E}Pr[e]\ p_{\rm succ}(\ketbra{\psi_e}),
    \end{eqnarray*}
    where $\pi_{\rm acc}$ is the projector onto the accepted measurement outcomes.
    
    The lower level is about the individual inputs $\ket{\phi_e}$ that are mapped to $\ket{\psi_e}$, which we find by decomposing the state on the higher level. Given a model of the single photon source we then use the Gram-Schmidt procedure to express these inputs in a spanning set of completely distinguishable photons. Note that this does not require purity of internal degrees of freedom, as one can use the same technique in Liouville space. However, we need to assume that there is no entanglement between the internal degrees of freedom and the position or polarization of the photon. Either way, we can represent the internal degrees of freedom as a single number being the element's index in the spanning set, which we call the distinguishability index. After the Gram-Schmidt procedure, the inputs are superpositions of states, each given by a combination in the six-folded cartesian product of the orthogonal set: 
    \begin{eqnarray}
        \ket{\phi_{\rm init}} &=& \sum_{\substack{\textbf{d}\in\{0,...,5\}^{\times 6}\\i<j\Rightarrow d_i\leq d_j}}\alpha_{\textbf{d}}\ket{n_{0,0,d_0},n_{1,0,d_1},n_{2,0,d_2},n_{3,0,d_3},n_{4,0,d_4},n_{5,0,d_5}},
    \end{eqnarray}
    where $n_{k,s,d}$ is the occupation number, i.e. the number of photons, in the spatial mode $k$ with polarization $s$ and the distinguishability index $d$. The weights $\alpha_{\textbf{d}}$ depend on the pairwise overlaps of the photons obtained by the Gram-Schmidt procedure. Hence, once one changes the pairwise overlaps one has to simulate the operation on the state from scratch unless one already simulates the operation on every constituent of the sum and uses the weights to rearrange the output. Our simulation uses this method to compute the fidelity of many overlaps in one run.
    
    

\subsection{Error models}
We consider three types of errors in our modelling -- distinguishability, higher-order terms and loss. Although our simulation method allows for a more general usage of these three types of error, we make some simplifications to reduce the dimensionality of the overall parameter space, which allows us to consider only five parameters in the end.

The first parameter is the \emph{overlap}, i.e., the inner product of the wave functions of the photons. With the overlap, we quantify the (partial) distinguishability\footnote{Another popular, albeit equivalent way to quantify the distinguishability is the \emph{HOM-dip visibility} \cite{HOM}, which is the square of the overlap for pure states.}; photons may be distinguishable due to degrees of freedom such as frequency and preparation time (which we consider constant during the GHZ state preparation). Note that for a different choice of gate encoding in which the polarization is an internal degree of freedom, one may have to consider the triad phase for three-photon interference \cite{Triad}. However, since the polarization is resolved by detection in our setup and the overlap of the wavefunctions is always real and positive, we do not have to consider this effect.  For simplicity, we assume that every pair of photons in the setup has the same overlap.

Higher-order terms in the state occur if two photons are emitted into the circuit instead of only one. The corresponding probability coincides with the (heralded) second-order correlation $g^{(2)}(\tau)$ function at $\tau=0$ and is our second parameter denoted as $g_2$. In the case of two-photon preparation, we replace the usual preparation map $\ket{0_a}\rightarrow \ket{1_a}$ at a mode $a$ by $\ket{0_a}\rightarrow \ket{2_a}$, i.e., we assume both photons are exactly in the same mode and indistinguishable. Theoretically, more than two photons could be emitted, but as the probability for terms of order higher than $2$ is usually very small, we omit to simulate these events. Photon pairs that are distinguishable, such as those generated via re-emission during an excitation process \cite{ollivier_hong-ou-mandel_2021}, will also contribute to gate errors. For the sake of simplicity, and due to the comparatively small values of the $g^{(2)}(0)$ we consider this to be an apt initial error model, even though it doesn’t perfectly capture the details of some solid state emitter sources. 
The last three parameters are connected to different types of loss that happen with some probability. We denote loss at preparation with $p_{L.Prep}$, loss at the beam splitters or the wave-plates with $p_{L.Ops}$ and loss at detection with $p_{L.Det}$. For single photon input states without higher order terms, loss only impacts the overall probability of success. However, the effect of loss also couples to contributions from higher-order photons.  We model loss as a map to a new spatial mode ($l$), i.e., if we denote with $n_{k,s,d}$ the occupation number in spatial mode $k$ with polarization $s$ and distinguishability value $d$, then loss acts as follows on the spatial mode $k$:
    \begin{eqnarray*}[rl]
        &\bigoplus_{d=0}^5\bigoplus_{s=0}^1\ket{n_{k,s,d}}\rightarrow
        \sum_{d'=0}^5\sum_{s'=0}^1\sqrt{\frac{n_{k,s',d'}}{N_k}}\bigoplus_{d=0}^5\bigoplus_{s=0}^1\ket{n_{k,s,d}-(\delta_{s,s'}\delta_{d,d'}),\ (\delta_{s,s'}\delta_{d,d'})_{l,s,d}},
    \end{eqnarray*}
    with 
    \begin{eqnarray*}[rl]
        N_k &= \sum_{d=0}^5\sum_{s=0}^1n_{k,s,d},\quad \quad
        \delta_{a,b}=\begin{cases}1,\ if\ a=b\\
        0,\ if\ a\neq b.\end{cases}
    \end{eqnarray*}
Intuitively, loss maps to a purification of the mixed state that one obtains if every photon has the same probability of getting lost. However, we do not consider the loss of multiple photons at the same step, as the probability of such coinciding loss is rather low.\\
Note that we use an error model for local GHZ state generation. Therefore, we omit modeling errors such as fiber losses that might be significant in communication settings. However, one could easily extend our model to include other types of losses.\\
There are two other types of errors that we do not consider. The first are dark counts, i.e. when a detector clicks without a photon incident. We expect this error to have a negligible effect on our circuit, as, with the advent of cyrogenic detection technology, dark count rates are typically small. Therefore, the probability that dark counts happen in the time window for expected photons is almost zero. The second type is imperfections of the gate implementations. 
Using our simulation method with Gaussian sampled rotation errors with a standard deviation of $1^{\circ}$, we found the effect on the fidelity is rather low, especially for low loss and $g^{(2)}$ setups.
Given the marginal effect and the significant computational overhead involved, we deemed rotational errors beyond the scope of this study.
\\[1em]

\subsection{Methods to evaluate the influence of imperfections}\label{sec:InfluenceDefs}
    One of the primary goals of this work is to evaluate the influence of imperfections on the fidelity and success probability. We use two quantities to evaluate the error parameters' influence on these measures. 
    
\begin{figure}
    \centering
    \label{fig:relIR}
    \includegraphics[width=0.75\textwidth]{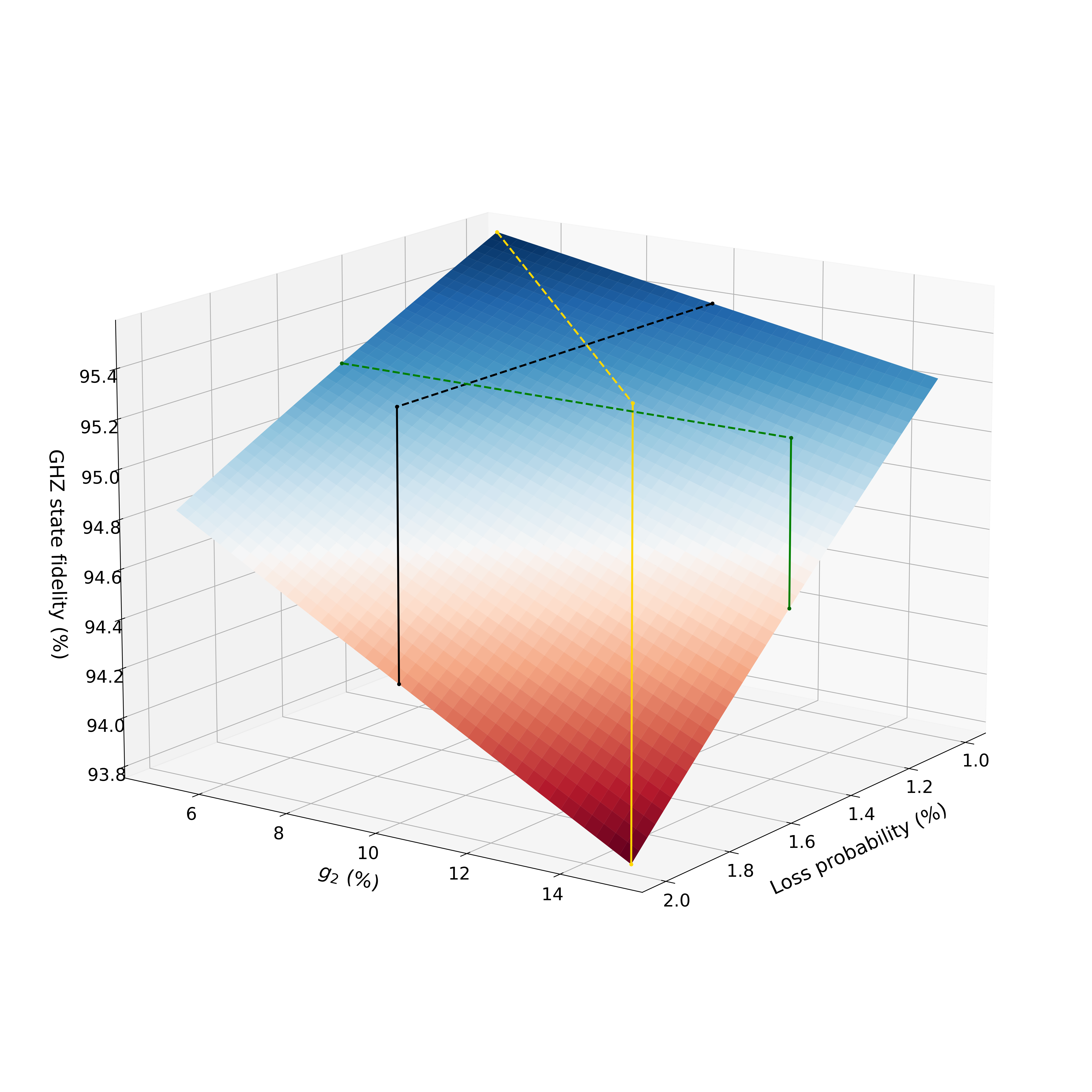}
    \caption{Visualization of the \emph{relative image range} using the GHZ state fidelity ($F_{GHZ}$) of the SPDC sources with an overlap of $99\%$ in a simplified loss model. In this figure, the relative image range $\Delta(p_L, F_{GHZ})$ with the loss probability is the ratio of the lengths of the solid black line and the solid yellow line. We would use the solid green line instead of the black line for the $\Delta(g_2, F_{GHZ})$.}
    \end{figure}

    The first quantity is the \emph{relative image range} $\Delta(p,m)$ of a parameter $p$ for a measure $m$. $\Delta(p,m)$ is the ratio of two differences -- the first difference is between the highest and lowest values of $m$ that can be obtained only by varying $p$ and using the default values for the other parameters. The second difference is between the maximum and minimum of $m$ for the whole parameter regime (cf. figure \ref{fig:relIR}). Intuitively, the relative image range measures the dependence of the measure on the parameter from an optimization perspective. It measures how much of the optimal outcome can be achieves by only varying the considered parameter. 
    More formally, we consider a measure $m$ to be a function from $N$ parameter ranges to $\mathbb{R}$. We define the \emph{relative image range} of the $i^{th}$ parameter for the measure $m$ as:
    \begin{eqnarray}\label{eq:IR}
        \Delta(p_i,m) = \frac{\max_{c_i\in P_i}(m(\bar{p}_1,...,c_i,...,\bar{p}_N)) - \min_{c_i\in P_i}(m(\bar{p}_1,...,c_i,...,\bar{p}_N))}{\max_{\mathbf{c}\in\mathbf{P}}(m(\mathbf{c})) - \min_{\mathbf{c}\in\mathbf{C}}(m(\mathbf{c}))},
    \end{eqnarray}
    where $\bar{p}_j$ denotes the default value, $P_j$ is the range of the parameter $c_j$ and $\mathbf{P} = P_1\times P_2\times ... \times P_N$.
    
    The second quantity is the \emph{correlation coefficient} $Corr(p,m)$ for an error parameter $p$ and a measure $m$. To estimate the correlation coefficient, we discretise the parameter regime and compute the measure for each point. After that, we compute the covariance $Cov(p,m)$ and the variances $Var(p)$ and $Var(m)$ and use 
    \begin{eqnarray}\label{eq:Cor}
        Corr(p,m) = \frac{Cov(p,m)}{\sqrt{Var(p)Var(m)}}. 
    \end{eqnarray}
    We assume a uniform distribution on the parameter regime for the covariance and variances. Under these conditions, we find:
    \begin{eqnarray*}
        Cov(p_i,m) &=& \sum_{c_1\in P^D_1}...\sum_{c_i\in P^D_i}...\sum_{c_N\in P^D_N}\frac{(c_i-\langle p_i\rangle)(m(c_1,...,c_i,...,c_N)-\langle m\rangle)}{\prod_{j=1}^N|P^D_j|}\\
        Var(p_i) &=& \sum_{c_i\in C^D_i}\frac{(c_i-\langle p_i\rangle)^2}{|P^D_i|}\\
        Var(m) &=&\frac{\sum_{\textbf{c}\in\textbf{P}^D}(m(\textbf{c})-\langle m\rangle)^2}{\prod_{j=1}^N|P^D_j|}
    \end{eqnarray*}
    where $C^D_j$ is the discretised parameter range for the $j^{th}$ parameter, $\textbf{P}^D = P^D_1\times P^D_2\times ... \times P^D_N$, 
    \begin{eqnarray*}
    \langle p_i\rangle=\frac{\sum_{c_i\in P^D_i}c_i}{|P^D_i|}\quad\text{and}\quad\quad \langle m\rangle=\frac{\sum_{\textbf{c}\in\textbf{P}^D}m(\textbf{c})}{\prod_{j=1}^N|P^D_j|}.
    \end{eqnarray*}
    In contrast to the relative image range, the correlation coefficient measures linear statistical dependence. On one side, this means that a high correlation coefficient indicates that increasing the parameter value will increase the value of the measure. On the other side, the potential for optimization can be lost in non-linearity. For example, $Corr(x, \sin(x))$ for $x\in[0,\pi]$ will be low while $\Delta(x, \sin(x))$ is $1$. Apart from quantifying linear dependence, the motivation to use the correlation coefficient is its role as the standard measure for correlation in statistics. With both quantities combined, we can investigate linear dependence and optimization potential. We find that for our use case, both quantities point to the same conclusion, which indicates approximately linear behavior and no hidden optimization potential.
    
    It is straightforward to compute the relative image range from the measure as it is apparent for which parameter value it takes its maximum and minimum value. However, for the correlation coefficient, there is a trade-off between the complexity of the error model and the resolution of the discretisation. For this reason, we simplify the modeling of the loss and introduce $p_L$ as the parameter for the context combination of the worst and best loss parameter combination, i.e., for every loss type we set
    \begin{eqnarray*}
        p_{T} = p_L \max_{c\in P_{T}}(c) + (1-p_L) \min_{c\in P_{T}}(c),
    \end{eqnarray*}
    where $p_T\in \{p_{L.Prep}, p_{L.Ops},p_{L.Det}\}$, $p_L\in[0,1]$ and $P_{T}$ is the associated parameter range.
    With this simplification, the overlap is varied in $0.25\%$ steps and for $g_2$ and $p_{L}$ we compute the measure for $201$ evenly distributed values in the corresponding ranges. Finally, note that the correlation coefficient can also be negative. Therefore, we will use correlation with $1-p$ for the loss and higher-order terms, since it holds that $Corr(1-p,m) = -Corr(p,m)$. 

\section{Results}
We investigate the fidelity, success probability, and the influence of the different error parameters on these measures in two realistic and one close-to-optimal parameter regime.
The two realistic parameter regimes represent two kinds of single photon sources: SPDC and solid-state sources. Currently, these two source types have different strengths and weaknesses. State-of-the-art SPDC sources can attain excellent two-photon indistinguishability (overlap), but are limited by comparatively large two-photon creation probabilities. In contrast, solid-state sources typically have significantly lower pairwise overlap but benefit from much reduced two-photon creation probability. For the close-to-optimal parameter regime, we assume almost no distinguishability and low loss probabilities. Table \ref{tab:parameter_regimes} shows all three parameter regimes and the measures we find for these regimes.

In the next section, we present the simulation results regarding the fidelity and the success probability for all three parameter regimes, and in Section \ref{sec:InfluenceAnalysis} we investigate the influence of the different error parameters on their values.

\begin{center}
    \resizebox{\textwidth}{!}{
    \begin{tabular}{l|r|r|r}
                                                   & SPDC \cite{Paesani2020,Kaneda:16,Weston:16,Graffitti:18}                     & Solid-state \cite{Kambs2018,cao2023photonic,Schweickert,ding2023highefficiency}                 & Close-to-optimal\\\hline
         Overlap ($ovl$) ($\%$):                   & 97.25 - 99.5 (99.0)   & 88.25 - 99.25 (94.0)      & 99 - 100 (99.5) \\
         Higher order ($g_2$) ($\%$):              & 1.0 - 2.0 (1.5)   & 0.0075 - 2.1 (0.25)      & 0 - 2 (1)\\
         Preparation loss ($p_{L.Prep.}$) ($\%$):  & 5.0 - 15.0 (10.0)   & 2.6 - 11.4 (7)      & 0 - 4 (2)\\
         Component loss ($p_{L.Ops.}$) ($\%$):     & 1.0 - 2.0 (1.5)   & 1.0 - 2.0 (1.5)      & 0 - 0.5 (0.25)\\
         Detection loss ($p_{L.Det.}$) ($\%$):     & 5.0 - 15.0 (10.0)   & 5.0 - 15.0 (10.0)      & 0 - 4 (2)\\\hline
         GHZ state Fidelity  ($\%$)                & 87.0 - 97.5 (94.8)   & 58.2 - 96.9 (77.3)      & 95.4 - 100.0 (97.8) \\
         Normalised Success probability ($\%$)     & 7.5 - 39.5 (17.8)   & 7.3 - 47.7 (19.5)      & 47.3 - 100.0 (68.6) \\
    \end{tabular}}
    \captionof{table}{Error parameters and measures in the different parameter regimes. The default (resp. expected) values are given in parentheses; if not specified differently we use the default values for the error parameter. For the close-to-optimal setting, we use the simplified error model. We normalized the success probability with respect to the one in the absence of errors (i.e., $1/32=3.125\%$).}
    \label{tab:parameter_regimes}
\end{center}

\subsection{Results of the simulation}\label{sec:baseline}
\begin{figure}[h!]
    \centering
    \includegraphics[width=\textwidth]{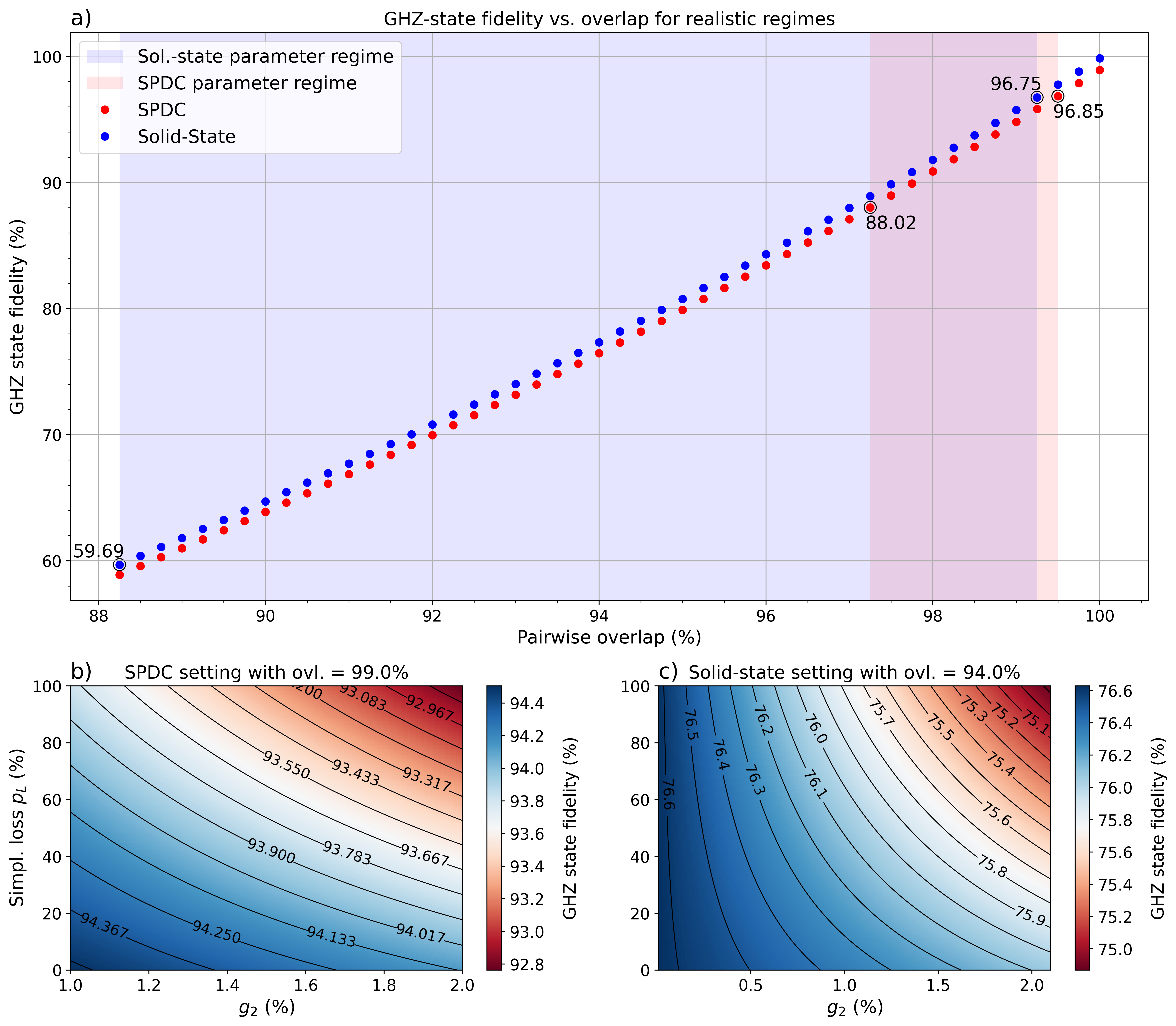}
    \caption{Subfigure a) shows the dependence of the fidelity on the overlap. The shaded areas indicate regimes realistic with the corresponding current technology. Subfigures b) and c) depict the dependence of the fidelity on loss and $g_2$, for SPDC and solid-state sources respectively in the simplified loss model.}
    \label{fig:fidVSovl}
\end{figure}
Our results indicate a high susceptibility of the circuit to experimental imperfections. Even for the best choice of parameters in the realistic regimes, the fidelities are significantly lower than in the close-to-optimal one, reflecting this susceptibility. The highest and lowest fidelity we estimate in the parameter regime we associate with the SPDC sources is $97.5\%$ and  $87.0\%$ respectively. The highest value is slightly larger than the best achievable fidelity in the solid-state parameter regime which is $96.9\%$. However, solid-state sources yield a broader range of fidelities, e.g., the lowest fidelity is $58.2\%$, while the fidelities with the expected values for the parameters shown in table \ref{tab:parameter_regimes}, are $94.8\%$ for the SPDC and $77.3\%$ for the solid-state sources. This difference implies that the fidelity is more susceptible to low overlap than to higher two-photon creation probability, as these two parameters pose the main difference between the two realistic parameter regimes. This implication is also demonstrated in figure \ref{fig:fidVSovl}, where one can see that there is more variation if one only varies the overlap (figure \ref{fig:fidVSovl}a) than with expected overlap and the other parameters being varied (figures \ref{fig:fidVSovl}b,c).

\begin{figure}[h!]
    \centering
    \resizebox{\textwidth}{!}{\includegraphics{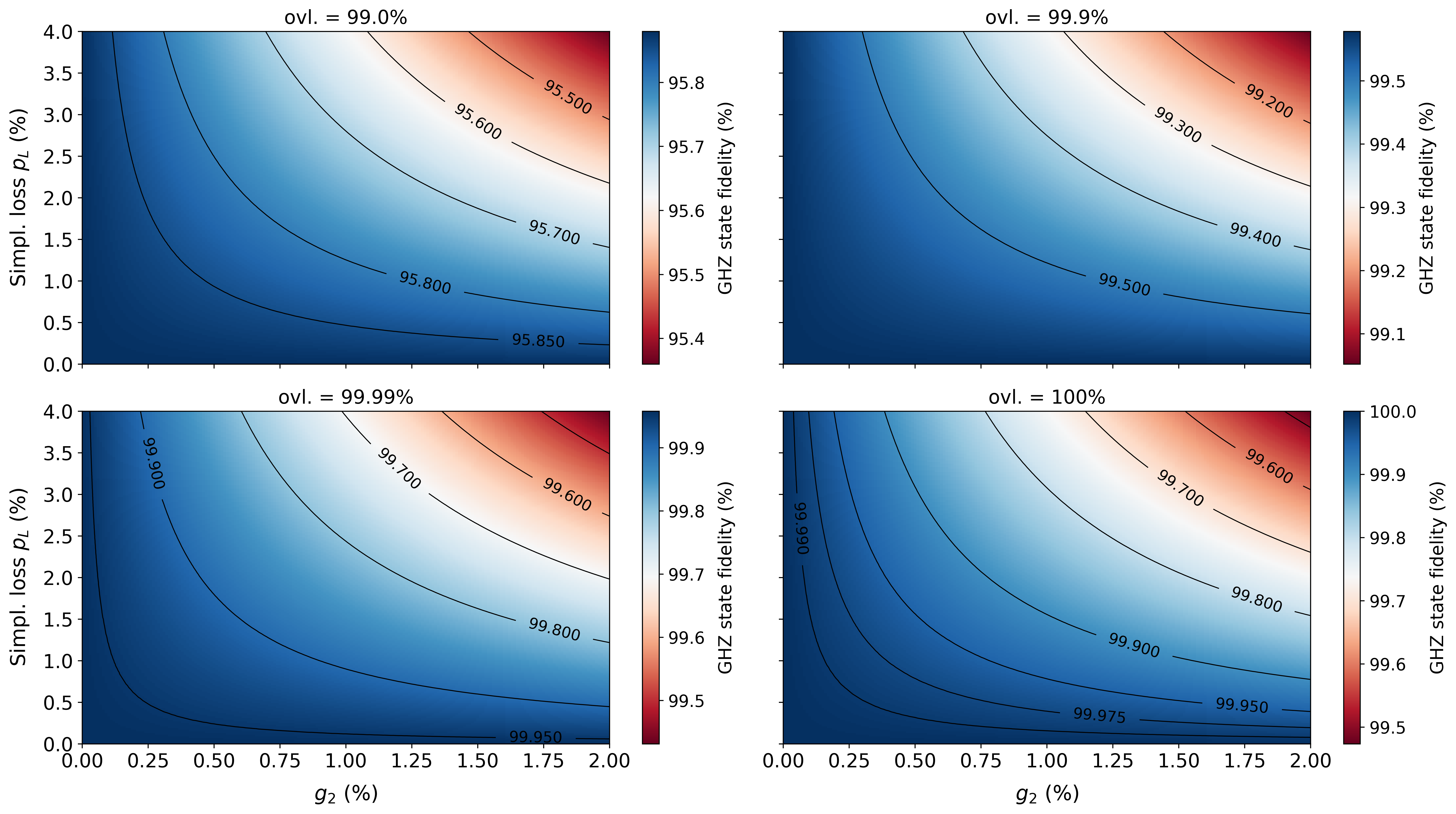}}
    \caption{The fidelity depicted for different overlaps versus $g_2$ and simplified loss in the close-to-optimal parameter regime.}
    \label{fig:P2vsPL_PS}
\end{figure}

Even in the small range between $99\%$ and $100\%$ overlap, the fidelity differs significantly: figure \ref{fig:P2vsPL_PS} shows the fidelity depending on the two-photon creation probability and simplified loss in the close-to-optimal parameter regime for different overlaps. Although the best fidelity in this regime is $100\%$, this value is only achieved if all parameters are optimal -- even with perfect overlap, one needs at least one of the other two parameters (simplified loss or two-photon creation probability) to be close to $0$ to reach a fidelity of $99.99\%$. However, with just $0.1\%$ less overlap, the fidelity is below $99.58\%$ even if the probability for loss and two-photon preparation is $0\%$.
\\

\begin{figure}[h]
    \centering
    \includegraphics[width=\textwidth]{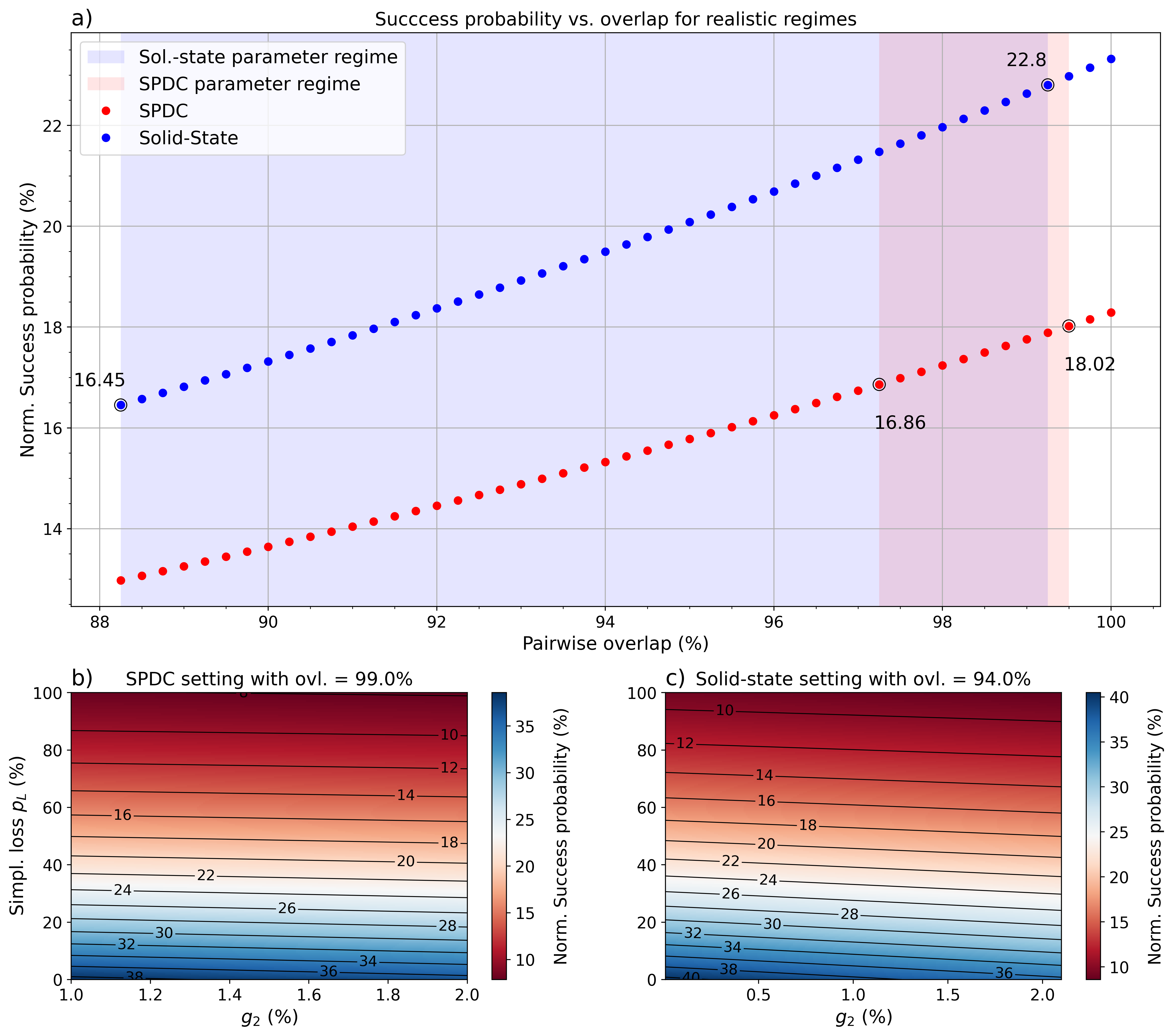}
    \caption{a) Success probability vs. overlap. The shaded areas indicate regimes realistic with current technology. Subfigures b) and c) depict the dependence of the success probability on loss and $g_2$ for SPDC and solid-state sources in the simplified loss model. The overlap has a lower influence on the success probability than on the fidelity. One can see that reducing the loss yields significantly higher success probabilities than achievable with the highest overlap and average loss. Here the higher-order terms play a minor role.}
    \label{fig:rateVsOvl}
\end{figure}
While the range of the fidelity of the SPDC sources includes the upper end of the range of solid-state sources, the situation is quite different when considering the normalized success probability (normalized with respect to the one in the absence of imperfections, which is $1/32$). The range of success probabilities for the SPDC parameter regime is contained in the one for the solid-state sources but is now at the lower end. Nevertheless, in both regimes, the success probability is low: The solid-state sources yield between $7.3\%$ and $47.7\%$ with an expected value of $19.5\%$, which is slightly better than the SPDC sources for which the success probability ranges between $7.5\%$ and $39.5\%$ with an expected value of $17.8\%$. A reason why the success probability is worse than the fidelity could be the lack of resilience against loss. In figure \ref{fig:rateVsOvl}, one can see that the fidelity is still relevant for the success probability (cf. figure \ref{fig:rateVsOvl}a), but seemingly not as important as the loss (cf. figure \ref{fig:rateVsOvl}b,c).

\begin{figure}[h]
    \centering
    \resizebox{\textwidth}{!}{\includegraphics{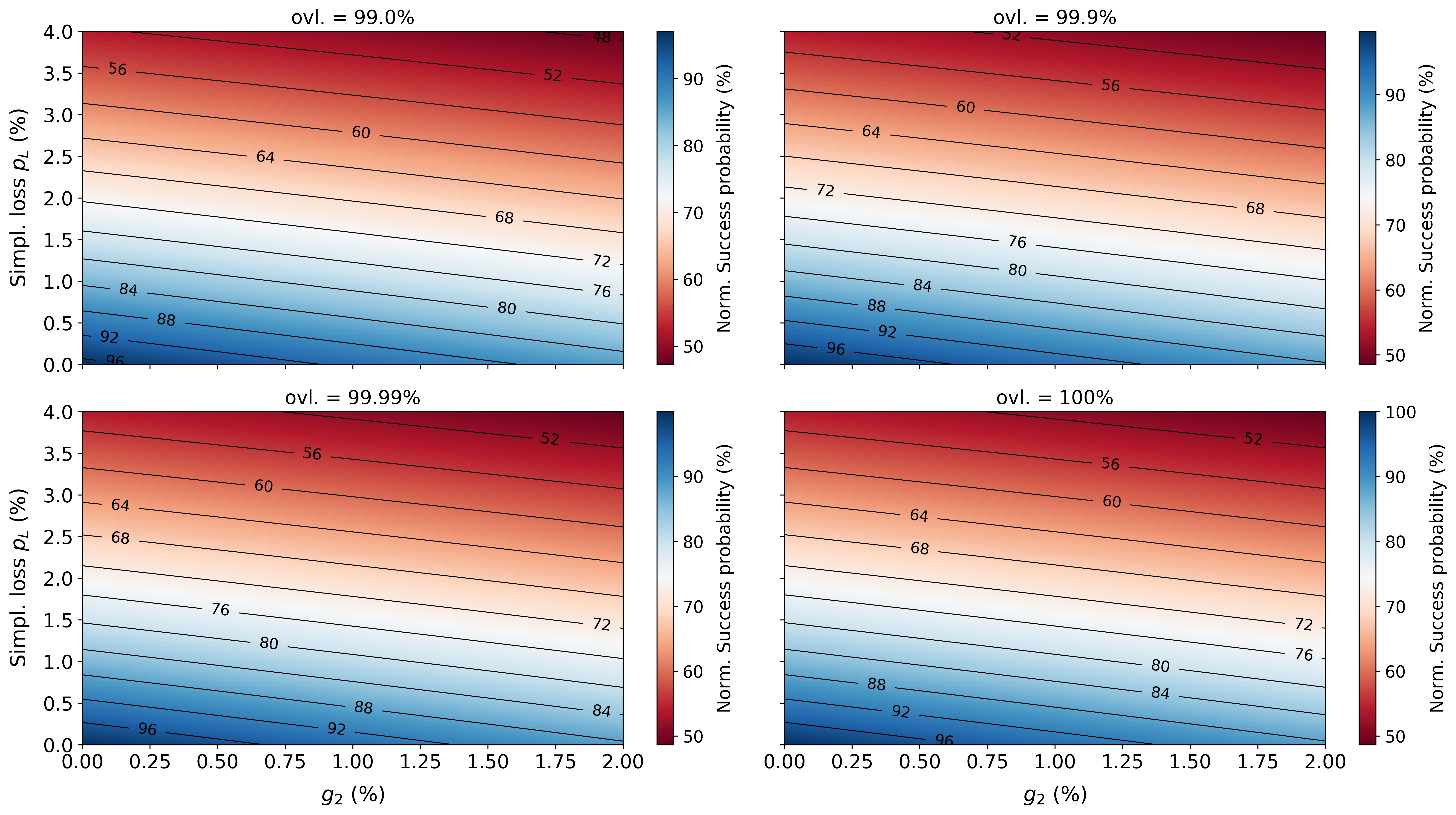}}
    \caption{The success probability depicted for different overlaps versus $g_2$ and simplified loss in the close-to-optimal parameter regime.}
    \label{fig:P2vsPL_Rate}
\end{figure}
In the close-to-optimal parameter regime, the circuit's success probability is also very susceptible to imperfections. Even with perfect overlap, most of the parameter regime maps to a normalized success probability that lays below $75\%$ (cf. figure \ref{fig:P2vsPL_Rate}) -- in fact, the high success probabilities above $96\%$ are barely visible in figure \ref{fig:P2vsPL_Rate}. This susceptibility is also confirmed by the lowest ($47.3\%$) and expected ($68.9\%$) normalized success probabilities in the parameter regime. Trivially, the highest normalized success probability is $100\%$ again.

\subsection{Influence analysis for the different types of imperfections}\label{sec:InfluenceAnalysis}

For both SPDC and solid-state sources, we find that the overlap is currently the crucial parameter influencing the fidelity. The relative image range of the overlap is $83.6\%$ for SPDC and $95.7\%$ for solid-state sources, and also the correlation coefficient of the overlap with the fidelity is large (SPDC: $99.0\%$; solid-state: $99.8\%$). In table \ref{tab:Eval} we show the evaluation quantities for the fidelity and in figure \ref{fig:fidVSovl}a) how the fidelity depends on the overlap.

Errors due to higher-order terms or loss have little influence on the fidelity in the realistic parameter regimes (cf. figure \ref{fig:fidVSovl}b,c). The fact that the influence of these parameters is significantly lower in the solid-state regime already hints at why loss and $g_2$ matter less.
The parameter regimes differ -- apart from the overlap -- mostly in the probability for higher-order terms. Post-selection causes the influence of the losses and $g_2$ on the fidelity to decrease if one of the parameters gets low enough. This effect becomes clearer if one considers $g_2=0$. In this case, the loss doesn't matter for the fidelity, as we filter out the state in the case of loss. Only with higher-order terms is it possible to accept the state if losses happen. The same argument works the other way around, i.e., in the absence of loss, higher-order terms have no influence. However, the loss probabilities are usually too large to be the limiting factor. Overall, we found that post-selection prevents a significant influence of loss and $g_2$ on the fidelity if one of them is low enough, which is the case in both parameter regimes. But especially in the solid-state parameter regime, the overlap is the only relevant parameter, as higher-order terms are vanishingly small.\\

\begin{table}[h]
    \centering
    \resizebox{\textwidth}{!}{
    \begin{tabular}{l|r|r|r|r|r|r}
    \multirow{3}{*}{} &
      \multicolumn{3}{c|}{Fidelity} &
      \multicolumn{3}{c}{Success prob.} \\
                            
                            & SPDC     & Sol.-state & Close-to-opt. &  SPDC     &  Sol.-state   &Close-to-opt.   \\\hline
$Corr(ovl, \cdot)$ $(\%)$   & $99.0$   & $99.8$       &$99.7$       & $4.3$  &   $22.3$       & $6.0$          \\
$Corr(1-g_2,\cdot)$ $(\%)$  & $8.0$   & $3.0$       &$5.0$       & $2.7$  &   $6.0$       & $18.4$         \\
$Corr(1-p_L,\cdot)$ $(\%)$  & $11.7$   & $1.6$       &$5.1$       & $98.1$  &   $95.3$       & $97.7$         \\\hline
$\Delta(ovl,\cdot)$ $(\%)$  & $83.6$   & $95.7$       &$88.8$       & $3.6$  &   $15.7$       & $3.9$          \\
$\Delta(g_2,\cdot)$ $(\%)$  & $6.7$   & $3.0$       &$5.6$       & $2.1$  &   $3.9$       & $14.7$         \\
$\Delta(p_L,\cdot)$ $(\%)$  & $9.8$   & $0.4$       &$5.6$       & $93.7$  &   $77.2$       & $80.0$      \\ 
    \end{tabular}}
    \caption{Evaluation of the influence using the relative image range $\Delta$ (cf. Eq. \ref{eq:IR}) and the correlation coefficient $Corr$ (cf. Eq. \ref{eq:Cor}) of the different error parameters only for the simplified loss. $p_L$ is the coefficient for the convex combination of maximum and minimum loss as described in the previous section.}
    \label{tab:Eval}
\end{table}
When considering the success probability, the overlap is still significant but is not the dominant parameter as it is for the fidelity (cf. figure \ref{fig:rateVsOvl}a)). The relative image range of the overlap is now only $3.6\%$ for SPDC sources and $15.7\%$ for solid-state sources -- so also for the success probability, the overlap has a larger influence on the solid-state sources than on the SPDC sources. 
The loss has an even bigger impact on the success probability -- the correlation between the simplified loss and the success probability for SPDC sources is almost as large as the correlation between the overlap and the fidelity. Even the individual loss parts are significant -- especially, preparation and detection loss have higher relative image ranges than $g_2$ or the overlap in both parameter regimes, only the component loss yields a lower image range than the overlap for solid-state sources. This difference in influence can also be seen in figures \ref{fig:rateVsOvl}b,c,
where the contour plots are almost horizontal which indicates a low dependency on the y-axis which shows the $g_2$ value.

\begin{table}[h]
    \centering
    \resizebox{\textwidth}{!}{
    \begin{tabular}{l|r|r|r|r|r|r}
    \multirow{3}{*}{} &
      \multicolumn{3}{c|}{Fidelity} &
      \multicolumn{3}{c}{Success prob.} \\
                                  & SPDC  & Sol.-state  &Close-to-opt.&  SPDC    &  Sol.-state& Close-to-opt.   \\\hline
$\Delta(p_{L.Prep},\cdot)$ $(\%)$  & $6.4$   & $0.2$       &$3.4$       & $36.3$  &   $27.5$       & $31.3$          \\
$\Delta(p_{L.Ops},\cdot)$  $(\%)$  & $1.8$   & $0.1$       &$3.1$       & $13.8$  &   $12.2$       & $30.0$         \\
$\Delta(p_{L.Det},\cdot)$   $(\%)$  & $1.6$   & $0.1$       &$0.8$       & $37.0$  &   $32.5$       & $31.7$         \\
    \end{tabular}}
    \caption{Evaluation of the influence for the different loss types using the relative image range $\Delta$ (cf. Eq. \ref{eq:IR}) on the fidelity and the success probability for all three parameter regimes.}
    \label{tab:EvalSpecLoss}
\end{table}

Another difference between the fidelity and success probability is how they are affected by the different kinds of loss, as shown in table \ref{tab:EvalSpecLoss}. For the fidelity, loss becomes more irrelevant during the process, i.e. loss at preparation is most significant and loss at detection is rather insignificant. This is probably because later loss is detected better by the post-selection as it is less likely to pair with a higher-order term. On the other hand, it makes no difference for the success probability, whether loss has happened in the preparation or detection process -- the relative image ranges for both cases are very close.

Many of the observations for the realistic parameter regimes hold in the close-to-optimal regime as well. The overlap influences the fidelity more (cf. figure \ref{fig:P2vsPL_PS}) and the success probability less (cf. figure \ref{fig:P2vsPL_Rate}). The loss is the most relevant parameter for the success probability and $g_2$ never has the most influence. Further, we again found that the loss becomes more insignificant for the fidelity throughout the process (cf. table \ref{tab:EvalSpecLoss}).
However, there are still differences. More specifically, higher-order terms are now more relevant than in the realistic parameter regimes. While in the latter, the correlation with $g_2$ and its relative image range are always lower than those of the overlap, in the close-to-optimal parameter regime the success probability depends more on $g_2$ than on the overlap. But also in general, $g_2$ is more relevant in this parameter regime, e.g. it has the same relative image range as the simplified loss for the fidelity and also the correlation coefficients are similar. 

\subsection*{Key findings}
\begin{itemize}
    \item The GHZ state fidelity of the outcome depends mainly on the overlap for all three parameter regimes. This is especially true for the solid-state parameter regime and the close-to-optimal one, where the low $g_2$ value prevents the influence of loss and higher-order terms.
    \item The success probability of the circuit depends mainly on the loss and is more independent of the overlap and the higher-order terms.
    \item For both measures mentioned above, the SPDC sources are more influenced by loss and higher-order terms than the solid-state sources which in turn are more influenced by the overlap.
    \item Only for the success probability in the close-to-optimal parameter regime, $g_2$ is more relevant than the overlap.
    \item In general, regarding the success probability, the solid-state sources perform better on average than the SPDC sources possibly due to the larger overlap independence of this measure. For the same reason, SPDC sources reach better fidelities.
\end{itemize}
\section{Discussion}

The use of multipartite entanglement is prevalent in photonic implementations of quantum communication and computation protocols. In this work, we investigate the influence of experimental imperfections on the generation of a well-known maximally entangled state, namely the GHZ state. We conducted large-scale simulations of a commonly-used circuit for GHZ state generation, using realistic parameters for SPDC and solid-state photonic sources, and we also investigated what happens in a close-to-optimal parameter regime. With the data obtained from our simulations, we analyzed the influence of different error parameters using the correlation coefficient as well as the newly-introduced relative image range. Our results indicate that present-day experimental setups are still too erroneous to implement sufficiently good GHZ states using the considered well-established generation method. To tackle this, we propose some steps towards generating better GHZ states. Improving the pairwise overlap should be the first priority to reach higher fidelities. Then, once we are in a high-overlap regime, we can choose between decreasing loss and higher-order terms, where decreased loss comes with the benefit of a significantly higher success probability.

Although the fidelity is amongst the most common measures for the quality of a state, it would be interesting to run our methods for different, more specialized measures. While it is widely believed that we need almost perfect GHZ states for implementing protocols using measurement-based quantum computing, for some cryptographic applications like conference key agreement \cite{CKArev}, imperfect GHZ states could also be used, combined with classical post-processing. For this purpose, a measure of non-classicality (e.g. contextuality) might be more interesting than the fidelity: One such candidate could be the success probability of the MBQC implementation of the OR-gate \cite{Raussendorf_2013}. In the context of quantum networks, it would also be interesting to combine this work with recent investigations on distributing GHZ states \cite{Avis,shimizu2024simple}.

Finally, besides varying the quality measure, one could also vary the method of generating the states and investigate different circuits. Our simulations allow us to understand better how to potentially improve the way we currently generate GHZ states, and thus gain better fidelities. However, as this might not be practically feasible in the near future, our work also aims to stimulate and promote research towards different preparation methods for photonic state generation.
\section{Code availability statement}
The code for the simulation is written in C++. Besides standard libraries, the Boost library \cite{BoostLibrary} was used. The code to generate the data used to compute the fidelities and success probabilities is openly available at the following URL/DOI: \href{https://doi.org/10.5281/zenodo.11518712}{10.5281/zenodo.11518712}.
\section{Data availability statement}
The data that support the findings of this study are openly available at the following URL/DOI: \href{https://doi.org/10.5281/zenodo.14168252}{10.5281/zenodo.14168252}.
\section{Acknowledgments}
This work was supported by the German Federal Ministry of Education and Research (BMBF project QPIC-1, No. 13N15870 and project DiNOQuant, No. 13N14921), by PiQ (a project of the DLR Quantum Computing Initiative, funded by the Federal Ministry for Economic Affairs and Climate Action), by Qompiler (funded by the German Federal Ministry for Economic Affairs and Climate Action), the DFG (via the Emmy Noether grant No.~418294583), the European Research Council (ERC, Starting Grant QUREP, NO. 851810), the Hector Fellow Academy and the Einstein Foundation Berlin (Einstein Research Unit on Quantum Devices).

\section{References}
\printbibliography[
heading=none]

@article{ollivier_hong-ou-mandel_2021,
	title = {Hong-{Ou}-{Mandel} {Interference} with {Imperfect} {Single} {Photon} {Sources}},
	volume = {126},
	issn = {0031-9007, 1079-7114},
	url = {https://link.aps.org/doi/10.1103/PhysRevLett.126.063602},
	language = {en},
	number = {6},
	urldate = {2024-10-02},
	journal = {Physical Review Letters},
	author = {Ollivier, H. and Thomas, S.E. and Wein, S. C. and De Buy Wenniger, I. Maillette and Coste, N. and Loredo, J. C. and Somaschi, N. and Harouri, A. and Lemaitre, A. and Sagnes, I. and Lanco, L. and Simon, C. and Anton, C. and Krebs, O. and Senellart, P.},
	month = feb,
	year = {2021},
	pages = {063602},
}

@article{Kambs2018,
  title = {Limitations on the indistinguishability of photons from remote solid state sources},
  volume = {20},
  ISSN = {1367-2630},
  url = {http://dx.doi.org/10.1088/1367-2630/aaea99},
  DOI = {10.1088/1367-2630/aaea99},
  number = {11},
  journal = {New Journal of Physics},
  publisher = {IOP Publishing},
  author = {Kambs,  Benjamin and Becher,  Christoph},
  year = {2018},
  month = nov,
  pages = {115003}
}

@article{omkar2022,
  title = {All-Photonic Architecture for Scalable Quantum Computing with Greenberger-Horne-Zeilinger States},
  author = {Omkar, Srikrishna and Lee, Seok-Hyung and Teo, Yong Siah and Lee, Seung-Woo and Jeong, Hyunseok},
  journal = {PRX Quantum},
  volume = {3},
  issue = {3},
  pages = {030309},
  numpages = {23},
  year = {2022},
  month = {7},
  publisher = {American Physical Society},
  doi = {10.1103/PRXQuantum.3.030309},
  url = {https://link.aps.org/doi/10.1103/PRXQuantum.3.030309}
}

@article{walther2004,
	title = {De {Broglie} wavelength of a non-local four-photon state},
	volume = {429},
	copyright = {2004 Macmillan Magazines Ltd.},
	issn = {1476-4687},
	url = {https://www.nature.com/articles/nature02552},
	doi = {10.1038/nature02552},
	language = {en},
	number = {6988},
	urldate = {2024-05-21},
	journal = {Nature},
	author = {Walther, Philip and Pan, Jian-Wei and Aspelmeyer, Markus and Ursin, Rupert and Gasparoni, Sara and Zeilinger, Anton},
	month = may,
	year = {2004},
	note = {Publisher: Nature Publishing Group},
	pages = {158--161},
}

@article{zhong2018,
	title = {12-{Photon} {Entanglement} and {Scalable} {Scattershot} {Boson} {Sampling} with {Optimal} {Entangled}-{Photon} {Pairs} from {Parametric} {Down}-{Conversion}},
	volume = {121},
	url = {https://link.aps.org/doi/10.1103/PhysRevLett.121.250505},
	doi = {10.1103/PhysRevLett.121.250505},
	number = {25},
	urldate = {2024-05-21},
	journal = {Physical Review Letters},
	author = {Zhong, Han-Sen and Li, Yuan and Li, Wei and Peng, Li-Chao and Su, Zu-En and Hu, Yi and He, Yu-Ming and Ding, Xing and Zhang, Weijun and Li, Hao and Zhang, Lu and Wang, Zhen and You, Lixing and Wang, Xi-Lin and Jiang, Xiao and Li, Li and Chen, Yu-Ao and Liu, Nai-Le and Lu, Chao-Yang and Pan, Jian-Wei},
	month = dec,
	year = {2018},
	note = {Publisher: American Physical Society},
	pages = {250505},
}

@article{bartolucci2023,
	title = {Fusion-based quantum computation},
	volume = {14},
	copyright = {2023 The Author(s)},
	issn = {2041-1723},
	url = {https://www.nature.com/articles/s41467-023-36493-1},
	doi = {10.1038/s41467-023-36493-1},
	language = {en},
	number = {1},
	urldate = {2024-05-21},
	journal = {Nature Communications},
	author = {Bartolucci, Sara and Birchall, Patrick and Bombín, Hector and Cable, Hugo and Dawson, Chris and Gimeno-Segovia, Mercedes and Johnston, Eric and Kieling, Konrad and Nickerson, Naomi and Pant, Mihir and Pastawski, Fernando and Rudolph, Terry and Sparrow, Chris},
	month = feb,
	year = {2023},
	note = {Publisher: Nature Publishing Group},
	pages = {912},
}

@article{varnava2008,
	title = {How {Good} {Must} {Single} {Photon} {Sources} and {Detectors} {Be} for {Efficient} {Linear} {Optical} {Quantum} {Computation}?},
	volume = {100},
	copyright = {http://link.aps.org/licenses/aps-default-license},
	issn = {0031-9007, 1079-7114},
	url = {https://link.aps.org/doi/10.1103/PhysRevLett.100.060502},
	doi = {10.1103/PhysRevLett.100.060502},
	language = {en},
	number = {6},
	urldate = {2024-05-21},
	journal = {Physical Review Letters},
	author = {Varnava, Michael and Browne, Daniel E. and Rudolph, Terry},
	month = feb,
	year = {2008},
	pages = {060502}
}

@article{schwartz2016,
	title = {Deterministic generation of a cluster state of entangled photons},
	volume = {354},
	url = {https://www.science.org/doi/10.1126/science.aah4758},
	doi = {10.1126/science.aah4758},
	number = {6311},
	urldate = {2024-05-21},
	journal = {Science},
	author = {Schwartz, I. and Cogan, D. and Schmidgall, E. R. and Don, Y. and Gantz, L. and Kenneth, O. and Lindner, N. H. and Gershoni, D.},
	month = oct,
	year = {2016},
	note = {Publisher: American Association for the Advancement of Science},
	pages = {434--437}
}

@article{istrati2020,
	title = {Sequential generation of linear cluster states from a single photon emitter},
	volume = {11},
	copyright = {2020 The Author(s)},
	issn = {2041-1723},
	url = {https://www.nature.com/articles/s41467-020-19341-4},
	doi = {10.1038/s41467-020-19341-4},
	language = {en},
	number = {1},
	urldate = {2024-05-21},
	journal = {Nature Communications},
	author = {Istrati, D. and Pilnyak, Y. and Loredo, J. C. and Antón, C. and Somaschi, N. and Hilaire, P. and Ollivier, H. and Esmann, M. and Cohen, L. and Vidro, L. and Millet, C. and Lemaître, A. and Sagnes, I. and Harouri, A. and Lanco, L. and Senellart, P. and Eisenberg, H. S.},
	month = oct,
	year = {2020},
	note = {Publisher: Nature Publishing Group},
	pages = {5501},
}

@article{wang2018,
	title = {18-{Qubit} {Entanglement} with {Six} {Photons}' {Three} {Degrees} of {Freedom}},
	volume = {120},
	url = {https://link.aps.org/doi/10.1103/PhysRevLett.120.260502},
	doi = {10.1103/PhysRevLett.120.260502},
	number = {26},
	urldate = {2024-05-21},
	journal = {Physical Review Letters},
	author = {Wang, Xi-Lin and Luo, Yi-Han and Huang, He-Liang and Chen, Ming-Cheng and Su, Zu-En and Liu, Chang and Chen, Chao and Li, Wei and Fang, Yu-Qiang and Jiang, Xiao and Zhang, Jun and Li, Li and Liu, Nai-Le and Lu, Chao-Yang and Pan, Jian-Wei},
	month = jun,
	year = {2018},
	note = {Publisher: American Physical Society},
	pages = {260502},
}

@article{thomas2024,
	title = {Fusion of deterministically generated photonic graph states},
	volume = {629},
	copyright = {2024 The Author(s)},
	issn = {1476-4687},
	url = {https://www.nature.com/articles/s41586-024-07357-5},
	doi = {10.1038/s41586-024-07357-5},
	language = {en},
	number = {8012},
	urldate = {2024-05-21},
	journal = {Nature},
	author = {Thomas, Philip and Ruscio, Leonardo and Morin, Olivier and Rempe, Gerhard},
	month = may,
	year = {2024},
	note = {Publisher: Nature Publishing Group},
	pages = {567--572},
}

@article{thomas2022,
	title = {Efficient generation of entangled multiphoton graph states from a single atom},
	volume = {608},
	copyright = {2022 The Author(s)},
	issn = {1476-4687},
	url = {https://www.nature.com/articles/s41586-022-04987-5},
	doi = {10.1038/s41586-022-04987-5},
	language = {en},
	number = {7924},
	urldate = {2024-05-21},
	journal = {Nature},
	author = {Thomas, Philip and Ruscio, Leonardo and Morin, Olivier and Rempe, Gerhard},
	month = aug,
	year = {2022},
	note = {Publisher: Nature Publishing Group},
	pages = {677--681},
}

@article{walther2005,
	title = {Experimental one-way quantum computing},
	volume = {434},
	copyright = {2005 Macmillan Magazines Ltd.},
	issn = {1476-4687},
	url = {https://www.nature.com/articles/nature03347},
	doi = {10.1038/nature03347},
    language = {en},
	number = {7030},
	urldate = {2024-05-21},
	journal = {Nature},
	author = {Walther, P. and Resch, K. J. and Rudolph, T. and Schenck, E. and Weinfurter, H. and Vedral, V. and Aspelmeyer, M. and Zeilinger, A.},
	month = mar,
	year = {2005},
	note = {Publisher: Nature Publishing Group},
	pages = {169--176},
}

@article{prevedel2007,
	title = {High-speed linear optics quantum computing using active feed-forward},
	volume = {445},
	copyright = {2006 Springer Nature Limited},
	issn = {1476-4687},
	url = {https://www.nature.com/articles/nature05346},
	doi = {10.1038/nature05346},
	language = {en},
	number = {7123},
	urldate = {2024-05-21},
	journal = {Nature},
	author = {Prevedel, Robert and Walther, Philip and Tiefenbacher, Felix and Böhi, Pascal and Kaltenbaek, Rainer and Jennewein, Thomas and Zeilinger, Anton},
	month = jan,
	year = {2007},
	note = {Publisher: Nature Publishing Group},
	pages = {65--69},
}

@article{zwerger2012,
	title = {Measurement-based quantum repeaters},
	volume = {85},
	copyright = {http://link.aps.org/licenses/aps-default-license},
	issn = {1050-2947, 1094-1622},
	url = {https://link.aps.org/doi/10.1103/PhysRevA.85.062326},
	doi = {10.1103/PhysRevA.85.062326},
	language = {en},
	number = {6},
	urldate = {2024-05-21},
	journal = {Physical Review A},
	author = {Zwerger, M. and Dür, W. and Briegel, H. J.},
	month = jun,
	year = {2012},
	pages = {062326},
}

@article{hilaire2023,
	title = {Near-deterministic hybrid generation of arbitrary photonic graph states using a single quantum emitter and linear optics},
	volume = {7},
	issn = {2521-327X},
	url = {http://arxiv.org/abs/2205.09750},
	doi = {10.22331/q-2023-04-27-992},
	language = {en},
	urldate = {2024-05-21},
	journal = {Quantum},
	author = {Hilaire, Paul and Vidro, Leonid and Eisenberg, Hagai S. and Economou, Sophia E.},
	month = apr,
	year = {2023},
	note = {arXiv:2205.09750 [quant-ph]},
	pages = {992},
}

@misc{chen2023heralded,
      title={Heralded three-photon entanglement from a single-photon source on a photonic chip}, 
      author={Si Chen and Li-Chao Peng and Yong-Peng Guo and Xue-Mei Gu and Xing Ding and Run-Ze Liu and Xiang You and Jian Qin and Yun-Fei Wang and Yu-Ming He and Jelmer J. Renema and Yong-Heng Huo and Hui Wang and Chao-Yang Lu and Jian-Wei Pan},
      year={2023},
      eprint={2307.02189},
      archivePrefix={arXiv},
      primaryClass={quant-ph}
}

@article{Thalacker_2021,
doi = {10.1088/1367-2630/ac1808},
url = {https://dx.doi.org/10.1088/1367-2630/ac1808},
year = {2021},
month = {08},
publisher = {IOP Publishing},
volume = {23},
number = {8},
pages = {083026},
author = {Christopher Thalacker and Frederik Hahn and Jarn de Jong and Anna Pappa and Stefanie Barz},
title = {Anonymous and secret communication in quantum networks},
journal = {New Journal of Physics}
}

@article{PhysRevA.59.1829,
  title = {Quantum secret sharing},
  author = {Hillery, Mark and Bu\ifmmode\check{z}\else\v{z}\fi{}ek, Vladim\'{\i}r and Berthiaume, Andr\'e},
  journal = {Phys. Rev. A},
  volume = {59},
  issue = {3},
  pages = {1829--1834},
  numpages = {0},
  year = {1999},
  month = {03},
  publisher = {American Physical Society},
  doi = {10.1103/PhysRevA.59.1829},
  url = {https://link.aps.org/doi/10.1103/PhysRevA.59.1829}
}

@article{Why...,
    author = {Rudolph, Terry},
    title = "{Why I am optimistic about the silicon-photonic route to quantum computing}",
    journal = {APL Photonics},
    volume = {2},
    number = {3},
    pages = {030901},
    year = {2017},
    month = {03},
    doi = {10.1063/1.4976737},
    url = {https://doi.org/10.1063/1.4976737},
    eprint = {https://pubs.aip.org/aip/app/article-pdf/doi/10.1063/1.4976737/14567582/030901\_1\_online.pdf},
}

@article{Raussendorf_2013,
	doi = {10.1103/physreva.88.022322},
	url = {https://doi.org/10.1103%2Fphysreva.88.022322},
	year = 2013,
	month = {08},
	publisher = {American Physical Society ({APS})},
	volume = {88},
	number = {2},
	author = {Robert Raussendorf},
	title = {Contextuality in measurement-based quantum computation},
	journal = {Physical Review A}
}

@misc{cao2023photonic,
      title={A photonic source of heralded GHZ states}, 
      author={H. Cao and L. M. Hansen and F. Giorgino and L. Carosini and P. Zahalka and F. Zilk and J. C. Loredo and P. Walther},
      year={2023},
      eprint={2308.05709},
      archivePrefix={arXiv},
      primaryClass={quant-ph}
}

@article{HOM,
  title = {Measurement of subpicosecond time intervals between two photons by interference},
  author = {Hong, C. K. and Ou, Z. Y. and Mandel, L.},
  journal = {Phys. Rev. Lett.},
  volume = {59},
  issue = {18},
  pages = {2044--2046},
  numpages = {0},
  year = {1987},
  month = {11},
  publisher = {American Physical Society},
  doi = {10.1103/PhysRevLett.59.2044},
  url = {https://link.aps.org/doi/10.1103/PhysRevLett.59.2044}
}

@article{Three_to_UQC,
  title = {From Three-Photon Greenberger-Horne-Zeilinger States to Ballistic Universal Quantum Computation},
  author = {Gimeno-Segovia, Mercedes and Shadbolt, Pete and Browne, Dan E. and Rudolph, Terry},
  journal = {Phys. Rev. Lett.},
  volume = {115},
  issue = {2},
  pages = {020502},
  numpages = {5},
  year = {2015},
  month = {07},
  publisher = {American Physical Society},
  doi = {10.1103/PhysRevLett.115.020502},
  url = {https://link.aps.org/doi/10.1103/PhysRevLett.115.020502}
}

@article{LOQCrev,
  title = {Linear optical quantum computing with photonic qubits},
  author = {Kok, Pieter and Munro, W. J. and Nemoto, Kae and Ralph, T. C. and Dowling, Jonathan P. and Milburn, G. J.},
  journal = {Rev. Mod. Phys.},
  volume = {79},
  issue = {1},
  pages = {135--174},
  numpages = {0},
  year = {2007},
  month = {01},
  publisher = {American Physical Society},
  doi = {10.1103/RevModPhys.79.135},
  url = {https://link.aps.org/doi/10.1103/RevModPhys.79.135}
}

@Article{Knill2001,
author={Knill, E.
and Laflamme, R.
and Milburn, G. J.},
title={A scheme for efficient quantum computation with linear optics},
journal={Nature},
year={2001},
month={01},
day={01},
volume={409},
number={6816},
pages={46-52},
issn={1476-4687},
doi={10.1038/35051009},
url={https://doi.org/10.1038/35051009}
}

@article{ACKA,
  title = {Anonymous Quantum Conference Key Agreement},
  author = {Hahn, Frederik and de Jong, Jarn and Pappa, Anna},
  journal = {PRX Quantum},
  volume = {1},
  issue = {2},
  pages = {020325},
  numpages = {12},
  year = {2020},
  month = {12},
  publisher = {American Physical Society},
  doi = {10.1103/PRXQuantum.1.020325},
  url = {https://link.aps.org/doi/10.1103/PRXQuantum.1.020325}
}

@article{CKArev,
author = {Murta, Gláucia and Grasselli, Federico and Kampermann, Hermann and Bruß, Dagmar},
title = {Quantum Conference Key Agreement: A Review},
journal = {Advanced Quantum Technologies},
volume = {3},
number = {11},
pages = {2000025},
doi = {https://doi.org/10.1002/qute.202000025},
url = {https://onlinelibrary.wiley.com/doi/abs/10.1002/qute.202000025},
eprint = {https://onlinelibrary.wiley.com/doi/pdf/10.1002/qute.202000025},
year = {2020}
}

@article{MBQC,
  title = {A One-Way Quantum Computer},
  author = {Raussendorf, Robert and Briegel, Hans J.},
  journal = {Phys. Rev. Lett.},
  volume = {86},
  issue = {22},
  pages = {5188--5191},
  numpages = {0},
  year = {2001},
  month = {05},
  publisher = {American Physical Society},
  doi = {10.1103/PhysRevLett.86.5188},
  url = {https://link.aps.org/doi/10.1103/PhysRevLett.86.5188}
}

@misc{jozsa2005introduction,
      title={An introduction to measurement based quantum computation}, 
      author={Richard Jozsa},
      year={2005},
      eprint={quant-ph/0508124},
      archivePrefix={arXiv},
      primaryClass={quant-ph}
}

@article{ExpGHZ,
  title = {Observation of Three-Photon Greenberger-Horne-Zeilinger Entanglement},
  author = {Bouwmeester, Dik and Pan, Jian-Wei and Daniell, Matthew and Weinfurter, Harald and Zeilinger, Anton},
  journal = {Phys. Rev. Lett.},
  volume = {82},
  issue = {7},
  pages = {1345--1349},
  numpages = {0},
  year = {1999},
  month = {02},
  publisher = {American Physical Society},
  doi = {10.1103/PhysRevLett.82.1345},
  url = {https://link.aps.org/doi/10.1103/PhysRevLett.82.1345}
}

@Inbook{GHZ,
author="Greenberger, Daniel M.
and Horne, Michael A.
and Zeilinger, Anton",
editor="Kafatos, Menas",
title="Going Beyond Bell's Theorem",
bookTitle="Bell's Theorem, Quantum Theory and Conceptions of the Universe",
year="1989",
publisher="Springer Netherlands",
address="Dordrecht",
pages="69--72",
isbn="978-94-017-0849-4",
doi="10.1007/978-94-017-0849-4_10",
url="https://doi.org/10.1007/978-94-017-0849-4_10"
}

@misc{BoostLibrary,
    author = "Boost",
    year   = 2024,
    title  = "{Boost C++ Libraries}",
    howpublished = "\url{http://www.boost.org/}"
}

@article{Schweickert,
    author = {Schweickert, Lucas and Jöns, Klaus D. and Zeuner, Katharina D. and Covre da Silva, Saimon Filipe and Huang, Huiying and Lettner, Thomas and Reindl, Marcus and Zichi, Julien and Trotta, Rinaldo and Rastelli, Armando and Zwiller, Val},
    title = "{On-demand generation of background-free single photons from a solid-state source}",
    journal = {Applied Physics Letters},
    volume = {112},
    number = {9},
    pages = {093106},
    year = {2018},
    month = {02},
    issn = {0003-6951},
    doi = {10.1063/1.5020038},
    url = {https://doi.org/10.1063/1.5020038},
    eprint = {https://pubs.aip.org/aip/apl/article-pdf/doi/10.1063/1.5020038/19756146/093106\_1\_online.pdf},
}

@article{parceval,
  doi = {10.22331/q-2023-02-21-931},
  url = {https://doi.org/10.22331/q-2023-02-21-931},
  title = {Perceval: {A} {S}oftware {P}latform for {D}iscrete {V}ariable {P}hotonic {Q}uantum {C}omputing},
  author = {Heurtel, Nicolas and Fyrillas, Andreas and Gliniasty, Gr{\'{e}}goire de and Le Bihan, Rapha{\"{e}}l and Malherbe, S{\'{e}}bastien and Pailhas, Marceau and Bertasi, Eric and Bourdoncle, Boris and Emeriau, Pierre-Emmanuel and Mezher, Rawad and Music, Luka and Belabas, Nadia and Valiron, Benoît and Senellart, Pascale and Mansfield, Shane and Senellart, Jean},
  journal = {{Quantum}},
  issn = {2521-327X},
  publisher = {{Verein zur F{\"{o}}rderung des Open Access Publizierens in den Quantenwissenschaften}},
  volume = {7},
  pages = {931},
  month = feb,
  year = {2023}
}

@article{Paesani2020,
	author = {Paesani, S. and Borghi, M. and Signorini, S. and Ma{\"\i}nos, A. and Pavesi, L. and Laing, A.},
	doi = {10.1038/s41467-020-16187-8},
	id = {Paesani2020},
	journal = {Nature Communications},
	number = {1},
	pages = {2505},
	title = {Near-ideal spontaneous photon sources in silicon quantum photonics},
	url = {https://doi.org/10.1038/s41467-020-16187-8},
	volume = {11},
	year = {2020}}

@article{Kaneda:16,
author = {Fumihiro Kaneda and Karina Garay-Palmett and Alfred B. U'Ren and Paul G. Kwiat},
journal = {Opt. Express},
number = {10},
pages = {10733--10747},
publisher = {Optica Publishing Group},
title = {Heralded single-photon source utilizing highly nondegenerate, spectrally factorable spontaneous parametric downconversion},
volume = {24},
month = may,
year = {2016},
url = {https://opg.optica.org/oe/abstract.cfm?URI=oe-24-10-10733},
doi = {10.1364/OE.24.010733}
}

@article{Weston:16,
author = {Morgan M. Weston and Helen M. Chrzanowski and Sabine Wollmann and Allen Boston and Joseph Ho and Lynden K. Shalm and Varun B. Verma and Michael S. Allman and Sae Woo Nam and Raj B. Patel and Sergei Slussarenko and Geoff J. Pryde},
journal = {Opt. Express},
number = {10},
pages = {10869--10879},
publisher = {Optica Publishing Group},
title = {Efficient and pure femtosecond-pulse-length source of polarization-entangled photons},
volume = {24},
month = may,
year = {2016},
url = {https://opg.optica.org/oe/abstract.cfm?URI=oe-24-10-10869},
doi = {10.1364/OE.24.010869}
}

@article{Graffitti:18,
author = {Francesco Graffitti and Peter Barrow and Massimiliano Proietti and Dmytro Kundys and Alessandro Fedrizzi},
journal = {Optica},
number = {5},
pages = {514--517},
publisher = {Optica Publishing Group},
title = {Independent high-purity photons created in domain-engineered crystals},
volume = {5},
month = may,
year = {2018},
url = {https://opg.optica.org/optica/abstract.cfm?URI=optica-5-5-514},
doi = {10.1364/OPTICA.5.000514}}

@article{Li:23,
author = {Chen-Long Li and Yao Fu and Wen-Bo Liu and Yuan-Mei Xie and Bing-Hong Li and Min-Gang Zhou and Hua-Lei Yin and Zeng-Bing Chen},
journal = {Opt. Lett.},
number = {5},
pages = {1244--1247},
publisher = {Optica Publishing Group},
title = {All-photonic quantum repeater for multipartite entanglement generation},
volume = {48},
month = {Mar},
year = {2023},
url = {https://opg.optica.org/ol/abstract.cfm?URI=ol-48-5-1244},
doi = {10.1364/OL.482287},
}

@misc{ding2023highefficiency,
      title={High-efficiency single-photon source above the loss-tolerant threshold for efficient linear optical quantum computing}, 
      author={Xing Ding and Yong-Peng Guo and Mo-Chi Xu and Run-Ze Liu and Geng-Yan Zou and Jun-Yi Zhao and Zhen-Xuan Ge and Qi-Hang Zhang and Hua-Liang Liu and Lin-Jun Wang and Ming-Cheng Chen and Hui Wang and Yu-Ming He and Yong-Heng Huo and Chao-Yang Lu and Jian-Wei Pan},
      year={2023},
      eprint={2311.08347},
      archivePrefix={arXiv},
      primaryClass={quant-ph}
}

@article{Hahn2019,
    title={Quantum network routing and local complementation},
    volume={5},
   ISSN={1520-8540},
   url={https://doi.org/10.1038/s41534-019-0191-6},
    DOI={10.1038/s41534-019-0191-6},
   number={76},
   journal={NPJ Quantum Info},
   author={Hahn, Frederik and Eisert, Jens and Anna Pappa},
   year={2019},
}

@ARTICLE{butterfly,
  author={Leung, Debbie and Oppenheim, Jonathan and Winter, Andreas},
  journal={IEEE Transactions on Information Theory}, 
  title={Quantum Network Communication—The Butterfly and Beyond}, 
  year={2010},
  volume={56},
  number={7},
  pages={3478-3490},
  doi={10.1109/TIT.2010.2048442},
}

@article{Kieling_2007,
   title={Minimal resources for linear optical one-way computing},
   volume={24},
   ISSN={1520-8540},
   url={http://dx.doi.org/10.1364/JOSAB.24.000184},
   DOI={10.1364/josab.24.000184},
   number={2},
   journal={Journal of the Optical Society of America B},
   publisher={Optica Publishing Group},
   author={Kieling, Konrad and Gross, David and Eisert, Jens},
   year={2007},
   month=jan, pages={184},
}

@misc{shimizu2024simple,
      title={Simple loss-tolerant protocol for GHZ-state distribution in a quantum network}, 
      author={Hikaru Shimizu and Wojciech Roga and David Elkouss and Masahiro Takeoka},
      year={2024},
      eprint={2404.19458},
      archivePrefix={arXiv},
      primaryClass={quant-ph}
}

@article{Avis,
  title = {Analysis of multipartite entanglement distribution using a central quantum-network node},
  author = {Avis, Guus and Rozpedek, Filip and Wehner, Stephanie},
  journal = {Phys. Rev. A},
  volume = {107},
  issue = {1},
  pages = {012609},
  numpages = {36},
  year = {2023},
  month = {Jan},
  publisher = {American Physical Society},
  doi = {10.1103/PhysRevA.107.012609},
  url = {https://link.aps.org/doi/10.1103/PhysRevA.107.012609}
}

@article{Triad,
  title = {Distinguishability and Many-Particle Interference},
  author = {Menssen, Adrian J. and Jones, Alex E. and Metcalf, Benjamin J. and Tichy, Malte C. and Barz, Stefanie and Kolthammer, W. Steven and Walmsley, Ian A.},
  journal = {Phys. Rev. Lett.},
  volume = {118},
  issue = {15},
  pages = {153603},
  numpages = {6},
  year = {2017},
  month = {Apr},
  publisher = {American Physical Society},
  doi = {10.1103/PhysRevLett.118.153603},
  url = {https://link.aps.org/doi/10.1103/PhysRevLett.118.153603}
}

@article{Maring24,
	author = {Maring, Nicolas and Fyrillas, Andreas and Pont, Mathias and Ivanov, Edouard and Stepanov, Petr and Margaria, Nico and Hease, William and Pishchagin, Anton and Lema{\^\i}tre, Aristide and Sagnes, Isabelle and Au, Thi Huong and Boissier, S{\'e}bastien and Bertasi, Eric and Baert, Aur{\'e}lien and Valdivia, Mario and Billard, Marie and Acar, Ozan and Brieussel, Alexandre and Mezher, Rawad and Somaschi, Niccolo},
	doi = {10.1038/s41566-024-01403-4},
	journal = {Nature Photonics},
	month = {03},
	pages = {1-7},
	title = {A versatile single-photon-based quantum computing platform},
	volume = {18},
	year = {2024}}

@article{Pont24,
	author = {Pont, Mathias and Corrielli, Giacomo and Fyrillas, Andreas and Agresti, Iris and Carvacho, Gonzalo and Maring, Nicolas and Emeriau, Pierre-Emmanuel and Ceccarelli, Francesco and Albiero, Ricardo and Dias Ferreira, Paulo Henrique and Somaschi, Niccolo and Senellart, Jean and Sagnes, Isabelle and Morassi, Martina and Lema{\^\i}tre, Aristide and Senellart, Pascale and Sciarrino, Fabio and Liscidini, Marco and Belabas, Nadia and Osellame, Roberto},
	doi = {10.1038/s41534-024-00830-z},
	journal = {npj Quantum Information},
	month = {05},
	title = {High-fidelity four-photon GHZ states on chip},
	volume = {10},
	year = {2024}}

@article{ExpCKA,
	author = {Massimiliano Proietti and Joseph Ho and Federico Grasselli and Peter Barrow and Mehul Malik and Alessandro Fedrizzi},
	doi = {10.1126/sciadv.abe0395},
	eprint = {https://www.science.org/doi/pdf/10.1126/sciadv.abe0395},
	journal = {Science Advances},
	number = {23},
	pages = {eabe0395},
	title = {Experimental quantum conference key agreement},
	url = {https://www.science.org/doi/abs/10.1126/sciadv.abe0395},
	volume = {7},
	year = {2021}}
%
%
%
%
%

\end{document}